\documentclass[preprint,showpacs,aps]{revtex4}
\usepackage{mathrsfs}
\begin{document}
\title{ Light Cone Structure near Null Infinity of the Kerr Metric}
\author{Shan Bai${}^{1,2}$, Zhoujian Cao${}^1$, Xuefei
Gong${}^{1,2,3}$, Yu Shang${}^{1,2}$, Xiaoning Wu$^1$ and
Y.K.Lau${}^{1,4}$} \affiliation{$^1$Institute of Applied
Mathematics, Academy of Mathematics and System Science, Chinese
Academy of Sciences, 55, Zhongguancun Donglu, Beijing, China,
100080} \affiliation{$^2$Graduate School of Chinese Academy of
Sciences, Beijing, China, 100080} \affiliation{$^3$Department of
Physics, Beijing Normal University,
 Beijing, China, 100875}
\affiliation{$^4$Albert-Einstein-Institut, Max-Planck-Institut fur
Gravitationsphysik, Am Muhlenberg 1, D-14476,  Golm, Germany }

\begin{abstract}

Motivated by our attempt to understand the question of angular
momentum of a relativistic rotating source carried away by
gravitational waves, in the asymptotic regime near future null
infinity of the Kerr metric, a family of null hypersurfaces
intersecting null infinity in shearfree (good) cuts are
constructed by means of asymptotic expansion of the eikonal
equation.
 The geometry of the null hypersurfaces as well as
the asymptotic structure of the Kerr metric near null infinity are
studied. To the lowest order in angular momentum, the Bondi-Sachs
form of the Kerr metric is  worked out. The Newman-Unti formalism is
then further developed, with which the Newman-Penrose constants of
the Kerr metric are computed and shown to be zero. Possible physical
implications of the vanishing of the Newman-Penrose constants of the
Kerr metric are also briefly discussed.
\end{abstract}
 \maketitle

\section{Introduction}

In general relativity, the angular momentum measured at future
null infinity (denoted by $\mathscr{I}^+$) of a  relativistic
rotating source is an enigmatic notion (see for instance
\cite{PRII}, chapter 9). For a generic weakly asymptotically
simple spacetime, the infinite dimensional supertranslation
symmetries  at $\mathscr{I}^+$ means that the definition of
angular momentum carried over from Minkowski space is not
canonically defined. The tidal force generated by the Weyl
curvature distorts an outgoing null hypersurface near
$\mathscr{I}^+$ in such a way that the spherical cut at which the
null hypersurface intersects $\mathscr{I}^+$ acquires non-trivial
shear structure  in the null direction tangential to the
hypersurface. This entails that, in general, it is not always
possible to find a family of shearfree cuts at $\mathscr{I}^+$
like that in a stationary spacetime and we are forced to treat,
subject to certain smoothness assumption, all cuts including those
with complicated shear structures on equal footing. Further,
 unlike the energy  of gravitational radiation,
 it does not seem to make sense to define a news function
describing the angular momentum carried away by gravitational
radiation.

This long standing problem emerges naturally also in a more
pragmatic context when we try to understand  the gravitational
waveform generated by a  relativistic rotating source near
$\mathscr{I}^+$, mainly motivated by numerical consideration (see
for instance \cite{winicour}). However, rather than following the
standard route of attempting to impose extra structure on
$\mathscr{I}^+$ to single out preferred cuts \cite{dain, rizzi}, the
generation of waveform calls for a better understanding of how a
measure of rotation (angular momentum), in some sense appropriately
defined, is encoded into the Bondi-Sachs  metric or its variants
\cite{NU, bartnik}. Underlying these  coordinates  is the
construction of a family of null hypersurfaces whose intersections
with $\mathscr{I}^+$ generate the Bondi time coordinate.

 As a preliminary step to seek further geometric
 insight into the problem, the present work purports to
 construct asymptotically, in the important example of a Kerr black hole in
 which angular momentum
   is well defined,
null hypersurfaces whose intersections with $\mathscr I^+$ generate
good cuts and try to see how angular momentum is encoded in the
geometry of the null hypersurfaces.
  The null hypersurfaces to be constructed here
are different from those considered by a number of authors
\cite{fletcher, hayward, bishop}. We will make further remarks
concerning this point in the next section.

The outline of the present article may be described as follows. In
Section 2, we will solve the eikonal equation for the Kerr metric
asymptotically near null infinity and then go on to construct a
family of null hypersurfaces intersecting $\mathscr I^+$. A
Newman-Penrose (NP) tetrad adapted to the null hypersurfaces is then
defined and used to study the geometry of the hypersurfaces as well
as the asymptotic structure of the Kerr metric near null infinity.
In section 3, the Bondi-Sachs form of the Kerr metric is worked out.
The Newman-Unti formalism will then be developed and the NP
constants for the Kerr metric will be calculated to be zero.
Possible physical implications of the vanishing of the NP constants
will also be briefly discussed. Throughout the present work, the
$(+,-,-,-)$ signature will be adopted for the spacetime metric and
we follow the NP notations in \cite{PRI}.

\section{Characteristc Structure  near Null Infinity}

\subsection{Eikonal Equation}

Let us start by looking at the explicit construction of a null
hypersurface in the  Kerr metric. To this end, locally, it is
sufficient to seek a smooth, real valued function $u$ such that the
eikonal equation
\begin{equation}
g^{ab}u_{,a}u_{,b}=0 \label{eikonal}
\end{equation}
is satisfied  where $g^{ab}$ is the contravariant form of the
spacetime metric. The most obvious solutions to the above eikonal
equation are obtained by means of separation of variables
(\cite{carter}, see also \cite{chandra}, chapter 7), as that used in
integrating the geodesic equations. The detailed geometry of
  these null hypersurfaces parameterized
 by a Carter constant, in particular the suspected singular behaviour
 along the symmetry axis,
is still under investigation and remains to be understood better.
See \cite{fletcher, hayward} in this context when the Carter
constant takes on the specific value $a^2$,  where $a$ has its
standard meaning in the Kerr metric.

In another work \cite{bishop}, Bondi-Sachs  coordinates for the Kerr
metric is constructed using solution to the eikonal equation
obtained by Pretorious and Isarel \cite{PI}. The detailed asymptotic
geometry of the Bondi-Sachs coordinates is yet to be analyzed.
Further, the metric coefficients of the Bondi-Sachs metric
 are given in terms of implicit functions and make them difficult to
  implement numerically.

In the present work, we shall put forward a new construction of  a
family of null hypersurfaces near $\mathscr I^+$, based upon which
Bondi-Sachs type coordinates may be constructed. One distinct
feature of our construction is that the intersections of these null
hypersurfaces with $\mathscr I^+$ generate good cuts, the existence
of which is characteristic of the asymptotic structure of a
stationary spacetime admitting $\mathscr I^+$.

To begin with, let $(t,r,\theta,\varphi)$ be the standard
Boyer-Lindquist coordinates of the Kerr metric. Consider first in
the flat space limit a light cone in Minkowski space described in
terms of the  oblate spheroid coordinates. The solution to the
eikonal equation in this case is given by
\begin{equation}
\label{light-min} u=t\,\pm\,\sqrt{r^2+a^2\sin^2\theta}
\end{equation} Apparently this solution cannot be obtained by
means of the conventional method of separable of variables. In the
case of the Kerr metric, to seek a solution of (\ref{eikonal})
without separation of variables and in the Minkowski space limit
degenerates into (\ref{light-min}) turns out to be quite difficult.
However, for our purpose, it is sufficient to seek an asymptotic
solution of (\ref{eikonal}) when $r$ is sufficiently large.

To see the way ahead, we first look at a light cone in the
Schwarzschild metric described by
\begin{eqnarray}
u=t-\ (\ r+2M\ln\frac{r-2M}{2M}\  )\label{schwar}
\end{eqnarray}
Asymptotically when $r$ is sufficiently large, (\ref{schwar})
becomes
\begin{eqnarray}
u=t-\left(r+2M\ln
\frac{r}{2M}-\frac{4M^2}{r}+\cdots\cdots\right)\label{schwar2}
\end{eqnarray}
The term $t-r-2M\ln \frac{r}{2M}$ survives in the asymptotic limit
and this guides us to adopt the following ansatz for the solution of
$u$ in (\ref{eikonal}) in the Kerr metric when $r$ is sufficiently
large,
\begin{eqnarray}
u=t-r-2M\ln
\frac{r}{2M}+\sum^{\infty}_{k=1}\frac{f_k}{r^k}\label{ansatz}
\end{eqnarray}
As we envisage the Bondi-Sachs type coordinates to be constructed
from the level sets of $u$ are axisymmetric,
 the
functions $f_k, k=1,2\cdots$ in (\ref{ansatz}) are then necessarily
functions of $\theta$ only and independent of $\varphi$.

The eikonal equation to be solved is given as
\begin{eqnarray}
r^2\left(1-\frac{2M}{r}+\frac{a^2}{r^2}\right)\left(\frac{\partial
u}{\partial r}\right)^2+\left(\frac{\partial
u}{\partial\theta}\right)^2=\frac{(r^2+a^2)^2}{r^2-2Mr+a^2}-a^2\sin^2\theta\label{eikonal2}
\end{eqnarray}

Substitute (\ref{ansatz}) into the eikonal equation above and solve
the eikonal equation order by order, we obtain
\begin{equation}
u=t-\left(r+2M\ln
\frac{r}{2M}-\frac{4M^2-\frac{1}{2}a^2\sin^2\theta}{r}-\frac{4M^3-Ma^2}{r^2}+
O(1/r^3)\right)\label{kerr}
\end{equation}
Inserting (\ref{ansatz}) into the eikonal equation in
(\ref{eikonal2}) enables us to solve  $f_k$ recursively. The ansatz
in (\ref{ansatz}) serves to determine uniquely the lowest order
terms $f_1$ and $f_2$. With $f_1$ and $f_2$ as initial conditions
for the algebraic process of repeated iterations of $f_k, k\ge 2$,
it may be checked that $f_{k+1}$ is determined uniquely by $f_r,
r=1,2\cdots k$. No freedom like, for instance, the existence of an
arbitrary, non-zero constant is allowed in each order. In principle
repeated iterations of $f_k$ generate terms of any desirable order
in the asymptotic expansion.
 However, as it occurs quite frequently in
  asymptotic expansion, the higher order terms
inevitably become  more complicated with the increase in order and
no regular pattern seems to be noticeable.

For a light cone in Minkowski spacetime  described in terms of
oblate spheroid coordinates, we have from (\ref{light-min}) that in
the asymptotic limit $r\rightarrow \infty$,
\begin{eqnarray}
u&=&t-(r+\frac{a^2\sin^2\theta}{2r}+\cdots\cdots)\label{minkowski}
\end{eqnarray}
This may also  be obtained from (\ref{kerr}) by taking the flat
space limit $M\rightarrow 0$, and thereby  provides a self
consistency check on the validity of (\ref{kerr}). Further, the flat
space and Schwarzschild limits of (\ref{kerr}) suggest that
  the constant $u$ hypersurfaces
constructed here are asymptotic parts of light cones emanating from
a  timelike world line.

\subsection{NP Tetrad and Asymptotic Structure near $\mathscr{I}^+$}

To study further  the geometry of the null hypersurfaces constructed
as well as  the asymptotic structure of the Kerr metric near future
null infinity, it will be helpful to define an NP tetrad adapted to
the constant $u$ null hypersurfaces. The null hypersurfaces
described in (\ref{kerr}) are outgoing. The dual ingoing null
hypersurfaces are given as
\begin{equation}
v=t+(r+2M\ln
\frac{r}{2M}-\frac{4M^2-\frac{1}{2}a^2\sin^2\theta}{r}-\frac{4M^3-Ma^2}{r^2}+\cdots\cdots)\label{kerrv}
\end{equation}
 Naturally, we choose two legs of the null tetrad to be parallel to the gradient
 vectors of
 $u$ and $v$.
In terms  of the Boyer-Lindquist coordinates,  the NP tetrad may
then be constructed as
\begin{eqnarray}
l_a&=&(du)_a=(1,-h_1,-h_2,0)\nonumber\\
n_a&=&\frac{1}{g^{00}-g^{11}\,h_1^{2}-g^{22}\,h_2^2}\,\,(1,\,h_1,\,h_2,\,0)\nonumber\\
m_a&=&\Big(\,g_{03}\,\frac{i}{\sin\theta}\,\sqrt{\frac{\rho^2}{2\Sigma^2}}\,,\,\,\,
-g_{11}\,\sqrt{\frac{-h_2^2}{2g_{11}\,h_2^2\,+\,2g_{22}\,h_1^2}}\,,
\nonumber\\
&&g_{22}\,\sqrt{\frac{-h_1^2}
{2g_{11}\,h_2^2\,+\,2g_{22}\,h_1^2}}\,,\quad
g_{33}\,\frac{i}{\sin\theta}\,\sqrt{\frac{\rho^2}{2\Sigma^2}}\,\Big)
\label{NP}
\end{eqnarray}
where $\Sigma^2=(r^2+a^2)^2-\Delta a^2\sin^2\theta$ and
$\rho^2=r^2+a^2\cos^2\theta$. $h_1$, $h_2$ are functions to be
determined by the solution (\ref{kerr}) and  may be solved
asymptotically order by order. With a view to compute the NP
constants for the Kerr metric later, we compute $h_1$, $h_2$ to
sufficiently high order so that the eikonal equation in
(\ref{eikonal})  is solved up to $1/ r^{7}$. The results are
\begin{eqnarray}
h_1&=&1+\frac{2M}{r}+\frac{4M^2-\frac{1}{2}a^2\sin^2\theta}{r^2}+\frac{8M^3-2Ma^2}{r^3}\nonumber
\\ &\ &+\frac{16M^4-8M^2a^2+\frac{3}{8}a^4\sin^4\theta}{r^4}
\nonumber \\
&\
&+\frac{32M^5-24M^3a^2+2Ma^4+\frac{1}{4}Ma^4\sin^4\theta}{r^5}\nonumber
\\ &\ &+\frac{64M^6-64M^4a^2+12M^2a^4-\frac{5}{16}a^6\sin^6\theta}{r^6}
\nonumber \\
&\
&+\frac{128M^7-160M^5a^2+48M^3a^4-2Ma^6-\frac{1}{2}Ma^6\sin^6\theta}{r^7}\nonumber
\\ &\ &+O(1/r^{8})
\nonumber \\
h_2&=&\frac{a^2\sin\theta\cos\theta}{r}-\frac{\frac{1}{2}a^4\sin^3\theta\cos\theta}{r^3}
\nonumber \\ &\ &-\frac{\frac{1}{4}Ma^4\sin^3\theta\cos\theta}{r^4}
+\frac{\frac{3}{8}a^6\sin^5\theta\cos\theta}{r^5}
+\frac{\frac{1}{2}Ma^6\sin^5\theta\cos\theta}{r^6}+O(1/r^{7}\label{h}
)
\end{eqnarray}
 The NP tetrad defined
in (\ref{NP}) is different from the standard Kinnersley tetrad as
$l_a$ is hypersurface forming. In the limit $a\rightarrow 0$, the
tetrad degenerates to the standard NP tetrad in the Schwarzschild
metric (\cite{chandra}, chapter 3).

The spin coefficients of the tetrad defined in (\ref{NP}) may
further be given  in the asymptotic limit $r\rightarrow \infty$ as
\begin{eqnarray}
\kappa&=&\kappa'=0 \nonumber \\
\rho&=&-\frac{1}{r}+\frac{a^2\sin^2\theta}{2r^3}+\frac{Ma^2\sin^2\theta}{2r^4}-\frac{3a^4\sin^4\theta}{8r^5}-\frac{7Ma^4\sin^4\theta}{8r^6}\nonumber
\\ &\
&+\frac{a^4\sin^4\theta(5a^2\sin^2\theta-16M^2)}{16r^7}+O(1/r^8)\nonumber\\
 \sigma&=&-\frac{3Ma^2\sin^2\theta}{2r^4}-\frac{5iMa^3\cos\theta\sin^2\theta}{r^5}+O(1/r^6)
\nonumber \\
\tau&=&-\frac{3iMa \sin\theta\ }
{\sqrt{2}r^3}+O(1/r^4)\nonumber \\
\epsilon&=&O({1}/{r^5}) \nonumber \\
\beta&=&\frac{\cot\theta}{2\sqrt{2}r}-\frac{1}{4\sqrt{2}\sin\theta}(-
a^2 \cos^3\theta+2 a^2 \cos\theta +6iMa\sin^2\theta)\frac{1}{r^3}
+O(1/r^4) \nonumber \\
\rho'&=&\frac{1}{2r}-\frac{M}{r^2}-\frac{a^2\sin^2\theta}{4r^3}+O(1/r^4)
\nonumber \\
\sigma'&=&\frac{3Ma^2\sin^2\theta}{4r^4}+O(1/r^5)
\nonumber \\
\tau'&=&-\frac{3iMa }{\sqrt{2}r^3}\sin\theta+O(1/r^4)
\nonumber \\
\epsilon'&=&-\frac{M}{2r^2}+ O(1/r^{4})\nonumber \\
\beta'&=&\frac{\cot\theta}{2\sqrt{2}r}-
\frac{1}{4\sqrt{2}\sin\theta}(-a^2 \cos^3\theta+2 a^2\cos\theta
+6iMa\sin^2\theta)\frac{1}{r^3}\nonumber\\ &&+O(1/r^4)\label{spin0}
\end{eqnarray}
The Weyl curvature components may also be given  asymptotically as
\begin{eqnarray*}
\Psi_0&=&\frac{3Ma^2 \sin^2
\theta}{r^5}+\frac{15iMa^3\sin^2\theta\cos\theta}{r^6}+O(1/r^{7})
\nonumber
\\
\Psi_1&=&\frac{3iMa \sin \theta}{\sqrt{2} r^4}-\frac{6\sqrt{2}Ma^2\sin\theta\cos\theta}{r^5}+O(1/r^{6}) \nonumber \\
\Psi_2&=&-\frac{M}{r^3}-\frac{3iMa\cos\theta}{r^4}+O(1/r^{5})\nonumber  \\
\Psi_3&=&-\frac{3iMa \sin \theta}{2 \sqrt{2} r^4}+O(/r^{5}) \nonumber \\
\Psi_4&=&\frac{3Ma^2 \sin^2 \theta}{4 r^5}+O(1/r^{6}) \nonumber
\end{eqnarray*}
From
$$g_{ab}=l_an_b+l_bn_a-m_a\bar m_b-\bar m_a m_b$$
together with (\ref{NP}) and (\ref{h}),
 the asymptotic form of the Kerr metric near future
null infinity may be worked out to be
\begin{eqnarray}
ds^2&=&\{1-\frac{2M}{r}+\frac{2Ma^2\cos^2\theta}{r^3}+O(1/r^4)\}du^2
 +\{1-\frac{a^2\sin^2\theta}{2r^2}+O(1/r^3)\}2du dr\nonumber\\
&+&\{\frac{a^2\sin\theta\cos\theta}{r}+O(1/r^2)\}2dud\theta
+\{\frac{2Ma\sin^2\theta}{r}+O(1/r^2)\}2dud\varphi+ O(1/r^4)\ dr^2\nonumber\\
&+&\{\frac{a^2\sin\theta\cos\theta}{r}+O(1/r^2)
\}2drd\theta+\{\frac{2Ma\sin^2\theta}{r}+O(1/r^2)
\}2drd\varphi\nonumber\\
&-&r^2\{1+\frac{a^2\cos^2\theta}{r^2}+O(1/r^3)
\}d\theta^2+\{\frac{2Ma^3\sin^3\theta\cos\theta}{r^2}
+O(1/r^3)\}2d\theta
d\varphi\nonumber\\
&-&r^2\{\sin^2\theta+\frac{a^2\sin^2\theta}{r^2}+O(1/r^3)
\}d\varphi^2\label{bs-light}
\end{eqnarray}
By the standard choice of conformal factor $\Omega=1/r$, the metric
in (\ref{bs-light}) may be conformally compactified. It may be
checked that the gradient of $\Omega$ at $\mathscr I^+$ is non-zero
and the second derivative of $\Omega$ vanishes at $\mathscr I^+$.
The structure of $\mathscr I^+$ for the Kerr metric is Minkowskian
 in the sense that the null generators are complete
and the topology is that of a light cone with its apex taken away
(i.e. topologically $S^2\times R$) \cite{geroch, NR}.  Further,  a
constant $u$ hypersurface intersects $\mathscr I^+$ in a unit two
sphere.

From (\ref{bs-light}), we observe that the zero and first order of
the metric coincide with that of Minkowski and Schwarzschild
respectively. Angular momentum appears in the terms of the order
$O(1/r^2)$. This is reminiscent of  the asymptotic behaviour of the
Kerr metric near spatial infinity.

 The non-null character of $dr$
in (\ref{bs-light}) means that the coordinates $(\theta,\varphi)$
are not constant along a null generator of a constant $u$
hypersurface.
 The
presence of angular momentum generates rotation of a constant $r$
spherical section during its motion along a constant $u$ null
hypersurface, taking along with it also the symmetry axis. This
suggests
 that the coordinates inherited from that of Boyer-Lindquist may
not be the natural one to work with near null infinity.
 This motivates us to further develop
the Newman-Unti (NU) formalism which describes a null generator of a
constant $u$ hypersurface in terms of its natural affine parameter
and the angular coordinates are those pulling back from null
infinity.

\section{NU Formalism and NP Constants}

 Suppose  $l^a=\left(\frac{\partial
}{\partial\lambda}\right)^a$ such that $\lambda$ is an affine
parameter of a null generator of a constant $u$ hypersurface. In
terms of the Boyer-Lindquist coordinates, we have from (\ref{NP})
and (\ref{h}) that
\begin{eqnarray}
l^a&=&\Big(1+\frac{2M}{r}+\frac{4M^2}{r^2}+\frac{8M^3-2Ma^2\cos^2\theta
}{r^3}+O(1/r^4),
\nonumber \\
&\ &\ \ 1+\frac{a^2\sin^2\theta}{2r^2}+\frac{Ma^2\sin^2\theta
}{r^3}-\frac{a^4(1+6\cos^2\theta-7\cos^4\theta)}{8r^4}-\frac{Ma^4(1-\cos^4\theta)}{2r^5}\nonumber
\\&\ &\ \ +\frac{a^6(1+5\cos^2\theta+3\cos^4\theta-9\cos^6\theta)-8M^2a^4\sin^4\theta}{16r^6}+O(1/r^7),\nonumber
\\&\ &\ \ \frac{a^2\sin\theta\cos\theta}{r^3}-\frac{a^4\sin\theta\cos\theta(1-\frac{1}{2}\sin^2\theta)}{r^5}-\frac{Ma^4\sin^3\theta\cos\theta}{4r^6}+O(1/r^7), \nonumber \\
&\ &\ \ \frac{2Ma}{r^3}+\frac{4M^2a}{r^4}+\frac{8M^3a-2Ma^3(1+\cos^2\theta)}{r^5}+O(1/r^6) \Big)\label{up}
\end{eqnarray}

From the definition of $l^a$ and (\ref{up}), we may obtain the
following coordinate transformations:
\begin{eqnarray}
r=&\lambda&-\frac{a^2\sin^2\tilde{\theta}}{2\lambda}-
\frac{Ma^2\sin^2\tilde{\theta}}{2\lambda^2}-\frac{a^4(1-6\cos^2\tilde{\theta}+5\cos^4\tilde{\theta})}{8\lambda^3}\nonumber
\\
&-&\frac{Ma^4(3-10\cos^2\tilde{\theta}+7\cos^4\tilde{\theta})}{8\lambda^4}\nonumber
\\&-&\frac{a^4\sin^2\tilde{\theta}[16M^2\sin^2\tilde{\theta}
+5a^2(1-14\cos^2\tilde{\theta}+21\cos^4\tilde{\theta})]}{80\lambda^5}+O(1/\lambda^{6})
\label{lambda}\\
\theta=&\tilde{\theta}&-\frac{a^2\sin\tilde{\theta}
\cos\tilde{\theta}}{2\lambda^2}+\frac{3a^4\sin\tilde{\theta}
\cos\tilde{\theta}\cos2\tilde{\theta}}{8\lambda^4}-
\frac{Ma^4\sin^3\tilde{\theta}\cos(\tilde{\theta})}{4\lambda^5}+O(1/\lambda^6)\nonumber
\\ \varphi=
&\tilde{\varphi}&-\frac{Ma}{\lambda^2}-\frac{4M^2a}{3\lambda^3}+O(1/\lambda^4)\,,
\nonumber
\end{eqnarray}
where $(\tilde \theta, \tilde\varphi)$ are the angular coordinates
pulled back from that of a cut at $\mathscr I^+$.

Before we move on, we digress at this point to work out the
Bondi-Sachs form of the Kerr metric, which is important for the
understanding of the characteristic structure  of the Kerr metric.
Define the luminosity parameter $\bar r$ in the standard way (see
for instance \cite{BMS}) by
\begin{eqnarray}
\partial_\lambda\, \bar{r}=-\rho\,\bar{r}. \label{lumi}
\end{eqnarray}
In view of (\ref{lambda}) and the explicit expression of $\rho$
given in terms of the Boyer-Lindquist coordinates in
(\ref{spin0}), it may be inferred that
\begin{eqnarray}
\rho=-\frac{1}{\lambda}-\frac{9M^2a^4\sin^4\tilde{\theta}}{20\lambda^7}+O(1/\lambda^8).
\label{rrho}
\end{eqnarray}
Integrating (\ref{lumi}) with the help of (\ref{rrho}), we then find
\begin{eqnarray}
\label{aff}
\bar{r}=\lambda-\frac{3M^2a^4\sin^4\tilde{\theta}}{40\lambda^5}+O(1/\lambda^6)
\end{eqnarray}
Substituting (\ref{lambda}) and (\ref{aff}) into the metric in
(\ref{bs-light}) or alternatively expressing the NP tetrad in
(\ref{NP}) in terms of the coordinates $(u,\bar
r,\tilde\theta,\tilde\varphi)$, we may  derive the Bondi-Sachs
form of the Kerr metric given as
\begin{eqnarray}
ds^2&=&\
\Big(1-\frac{2M}{\bar{r}}+\frac{Ma^2(2\cos^2\tilde{\theta}-\sin^2\tilde{\theta})}{\bar{r}^3}
+O(1/\bar{r}^4)\Big)du^2 \nonumber \\&\
&+\Big(2-\frac{15a^4\sin^2\tilde{\theta}(5+8\cos2\tilde{\theta}-\cos^2\tilde{\theta})}
{4\bar{r}^4}+O(1/\bar{r}^5)\Big)dud\bar{r} \nonumber \\ &\
&-\Big(\frac{6Ma^2\sin\tilde{\theta}\cos\tilde{\theta}}{\bar{r}^2}+O(1/\bar{r}^3)\Big)dud\tilde{\theta}
+\Big(\frac{4Ma\sin^2\tilde{\theta}}{\bar{r}}+O(1/\bar{r}^2)\Big)dud\tilde{\varphi}\nonumber
\\ &\
&-\Big(\bar{r}^2-\frac{Ma^2\sin^2\tilde{\theta}}{\bar{r}}+O(1/\bar{r}^2)\Big)d\tilde{\theta}^2
-\Big(\frac{12M^2a^3\sin^3\tilde{\theta}\cos\tilde{\theta}}{\bar{r}^3}+O(1/\bar{r}^4)\Big)d\tilde{\theta}d\tilde{\varphi}
\nonumber \\ &\
&-\Big(\bar{r}^2\sin^2\tilde{\theta}+\frac{Ma^2\sin^4\tilde{\theta}}{\bar{r}}
+O(1/\bar{r}^2)\Big)d\tilde{\varphi}^2. \nonumber
\end{eqnarray}
In principle,  with more involved calculations, higher order terms
of the Bondi-Sachs form of the metric may be generated using the
same algorithm.

Now we return to our discussion of the NU formalism for the Kerr
metric, and we shall adopt the coordinates
$\{u,\lambda,\tilde{\theta},\tilde{\phi}\}$ again in our subsequent
discussions. The NP tetrad given in (\ref{NP}) and (\ref{h}) are not
parallelly transported along a generator of a  constant $u$
hypersurface. This manifests in the non-zeroness of $\tau'$ and the
imaginary part of $\epsilon$ in (\ref{spin0}). In the next step,
with $l^a$ kept fixed, we shall rotate the NP tetrad defined in
(\ref{NP}) into one which is parallelly transported along a null
generator of a constant $u$ hypersurface, again in an order by order
fashion. To this end, we first rotate $m^a$ by a phase angle, i.e.
$m^a\rightarrow e^{i\chi}m^a$ where $\chi$ is a real valued function
of $\lambda,\tilde\theta,\tilde\phi$. The spin coefficient
$\epsilon$ transforms accordingly as
\begin{eqnarray}
\epsilon\rightarrow \epsilon+\frac{1}{2}i\,l^a\nabla_a\,\chi
\nonumber
\end{eqnarray}
For $m^a$ to be parallelly transported along a null generator with
tangent vector $l^a$, the vanishing of $\epsilon$ requires
\begin{equation}
\label{free} \chi=\chi_0(\tilde\theta,
\tilde\varphi)+O(1/\lambda^{4}), \end{equation} where $\chi_0$ is an
arbitrary, real valued function defined on a unit two sphere and it
signifies the $SO(2)$ degrees of freedom in the definition of $m^a$
and $\bar m^a$ with $l^a, n^a$ fixed. By further stipulating that
asymptotically  the angular part of $m^a$ should take the form
$\frac{1}{\sqrt {2}r}\left(\partial_{\tilde\theta}
+i\csc\tilde\theta\,\partial_{\tilde\varphi}\right)$, we may choose
$\chi_0$ to be zero and
  conclude from (\ref{free}) that
 \begin{equation}
\label{free1} \chi=O(\lambda^{-4}). \end{equation}
 This will be
sufficient for the calculation of the NP constants to be considered
in a moment.

For $n^a$ to be parallelly transported, we need to perform the null
rotation
\begin{eqnarray}
l^a\rightarrow l^a,\, m^a\rightarrow m^a+b\, l^a, \ n^a\rightarrow
n^a+\bar{b}\,m^a+b\,\bar m^a+b\bar{b}\,l^a\label{rotate}
\end{eqnarray}
where $b$ is a complex valued function of $\lambda, \tilde \theta,
\tilde\varphi$ and $\bar b$ denotes its complex conjugation. Subject
to (\ref{rotate}), $\epsilon$ remains unchanged due to $\kappa=0$,
while $\tau'$ transforms as
\begin{eqnarray}
\tau'\rightarrow\tau'-2\bar b\,\epsilon-l^a\nabla_a\bar
b.\label{tau}
\end{eqnarray}
With the help of (\ref{spin0}) and (\ref{free1}), from (\ref{tau})
we may infer that, for $\tau'$ to vanish, we require
\begin{eqnarray}
b=\frac{3iMa\sin\tilde\theta}{2\sqrt{2}\lambda^2}-\frac{Ma^2\sin\tilde{\theta}\cos\tilde{\theta}}{\sqrt{2}\lambda^3}+O(1/\lambda^{4})\,.\label{b}
\end{eqnarray}
The constant of integration is set to zero in (\ref{b}) so that
asymptotically the component of $m^a$ in the $\frac{\partial}
{\partial \lambda}$ direction starts from the order of
$O(1/\lambda)$.

For the parallelly transported NP tetrad on a constant $u$
hypersurface, the corresponding spin coefficients and the peeling
off behaviour of the Weyl curvature components may then be worked
out to be
\begin{eqnarray}
\kappa&=&\epsilon=\kappa'=\tau'=0 \nonumber \\
\rho&=&-\frac{1}{\lambda}-\frac{9M^2a^4\sin^4\tilde{\theta}}{20\lambda^7}+O(1/\lambda^8) \nonumber \\
\sigma
&=&-\frac{3Ma^2\sin^2\tilde\theta}{2\lambda^4}-\frac{5iMa^3\sin^2\tilde\theta\cos\tilde\theta
}{r^5}+O(1/\lambda^{6})\nonumber
\\
\tau&=&-\frac{3iMa\sin\tilde\theta}{2\sqrt{2}\lambda^3}
+\frac{2\sqrt{2}Ma^2\sin\tilde\theta\cos\tilde\theta}{\lambda^4}+O(1/\lambda^{5})\nonumber
\\
\beta&=&\frac{\cot\tilde\theta}{2\sqrt{2}\lambda}
-\frac{3iMa\sin\tilde\theta}{2\sqrt{2}\lambda^3}+O(1/\lambda^{4})\nonumber
\\
\rho'&=&\frac{1}{2\lambda}-\frac{M}{\lambda^2}-\frac{3iMa\cos\tilde\theta}{2\lambda^3}
+O(1/\lambda^{4})\nonumber
\\
\sigma'&=&\frac{Ma^2\sin^2\tilde\theta}{4\lambda^4}+O(1/\lambda^5)\nonumber  \\
\epsilon'&=&-\frac{M}{2\lambda^2}-\frac{3iMa\cos\tilde\theta}{4\lambda^3}+O(1/\lambda^{4})\nonumber
\\
\beta'&=&\frac{\cot\tilde\theta}{2\sqrt{2}\lambda}+O(1/
\lambda^{4})\label{spin}
\end{eqnarray}
and
\begin{eqnarray}
\Psi_0&=&\frac{3Ma^2\sin^2\tilde\theta}{\lambda^5}
+\frac{15iMa^3\sin^2\tilde\theta\cos\tilde\theta}{\lambda^6}+O(1/\lambda^{7})\nonumber
\\
\Psi_1&=&\frac{3iMa\sin\tilde\theta}{\sqrt{2}\lambda^4}
-\frac{6\sqrt{2}Ma^2\sin\tilde\theta\cos\tilde\theta}{\lambda^5}+O(1/\lambda^{6})\nonumber
\\
\Psi_2&=&-\frac{M}{\lambda^3}-\frac{3iMa\cos\tilde\theta}{\lambda^4}+O(1/\lambda^{5})
\nonumber
\\
\Psi_3&=&-\frac{3iMa\sin\tilde\theta}{2\sqrt{2}\lambda^4}+O(1/\lambda^{5})\nonumber
\\
\Psi_4&=&\frac{3Ma^2\sin^2\tilde\theta}{4\lambda^5}+O(1/\lambda^{6})\,.\label{weylnu}
\end{eqnarray}
We may see from the spin coefficient $\rho$ in (\ref{spin}) that,
once the Boyer-Lindquist coordinates is chosen, the scaling freedom
for the affine parameter is also determined. From the asymptotic
behaviour of the spin coefficient $\sigma$, it may also be seen that
the asymptotic shear defined by $\Omega^{-2}\sigma$ responsible for
the news vanishes on the unit sphere at which a constant u
hypersurface intersects $\mathscr I^+$. This existence of this kind
of good cuts is characteristic of the asymptotic structure of a
stationary, weakly asymptotically simple spacetime in which the
gravitational radiation field defined by $\Omega^{-1}\Psi_4$
vanishes \cite{NP}.

Define $\hat \tau=\Omega^{-3}\tau$ and
$\hat\Psi_{1}=\Omega^{-4}\Psi_1$. The angular momentum of the Kerr
metric may be expressed as
\begin{eqnarray}
Ma&=&-\frac{\sqrt 2}{\,\,\,3i\pi^2\,}\int \,\hat\tau\,d\hat S\label{tau-ang}\\
&=&\,\,\,\,\frac{\sqrt 2}{\,\,\,3i\pi^2\,}\int \,\hat\Psi_1\,d\hat
S,\nonumber
\end{eqnarray}
where the integration is over a unit two sphere at $\mathscr I^+$.
Using the NP equations and the explicit expressions of the spin
coefficients given in (\ref{spin}), it may be deduced that the above
angular momentum expressions are all special cases of the the
linkage expression \cite{link} (or equivalently the Komar integral)
written in terms of the NP tetrad chosen here.
  From (\ref{tau-ang}), we also see that  the angular
momentum is a measure of non-integrability of the timelike two plane
spanned by the null vectors $l^a, n^a$ (see also \cite{hay} in this
connection).

\subsection{NP constants of the Kerr Metric}

With the NU framework we have developed and the calculations we have
done on various quantities, we are now in a position to further
compute the NP constants for the Kerr metric.

 Consider the NP constants \cite{NP} defined as
  \begin{equation}
  \label{npconst}
 G_m=\int_0^{2\pi}\int_0^{\pi}  {} _2\bar{Y}_{2m}\,\Psi_0^1 \,\sin\tilde{\theta}\, d\tilde{\theta}
 d\tilde{\phi},\,\, m=0,\pm 1,\pm 2
 \end{equation}
where ${} _2Y_{2m}$ are the spin weight 2 spherical harmonics.
$\Psi_0^1$ is defined by the asymptotic expansion of $\Psi_0$ as
$$\Psi_0=\frac{\Psi_0^0}{\lambda^5}
+\frac{\Psi_0^1}{\lambda^6}+O(1/\lambda^{7}).$$
 Axisymmetry of the
Kerr metric means that $G_{\pm 1}$ and $G_{\pm 2}$ vanish trivially.
With $$\Psi_0^1=15iMa^3\sin^2\tilde\theta\cos\tilde\theta$$
according to (\ref{weylnu}), it may be calculated easily from
(\ref{npconst}) that $G_0=0$  and therefore all NP constants vanish
in a Kerr metric.

 Alternatively, with the definitions of multipole moments defined in \cite{JN},
 we may  also work out from (\ref{weylnu}) the explicit expressions
 for
 the monopole ($\cal M$), dipole ($\cal D$) and quadruple ($\cal Q$) moments
  and may be given respectively
 as
\begin{eqnarray*}
\cal M&=&
-\frac{1}{4}\int_0^\pi(\Psi_2^0+\bar{\Psi}_2^0)\sin\tilde{\theta}d\tilde{\theta}=M\\
\cal D&=&
-\frac{1}{2\sqrt{2}}\int_0^\pi\Psi_1^0\,P_1^1(\cos\tilde{\theta})\sin\tilde{\theta}d\tilde{\theta}=iMa\\
\cal Q&=&
-\frac{5}{24}\int_0^\pi\Psi_0^0\,P_2^2(\cos\tilde{\theta})\sin\tilde{\theta}
d\tilde{\theta}=-2Ma^2\,,
\end{eqnarray*}
where $P_l{}^m$ are the standard Legendre polynomials. The  NP
constant $G_0$  may also be calculated from the formula \cite{NP}
\begin{eqnarray}
G_0=2\sqrt{30\pi}(2\mathcal{D}^{2}-\mathcal{M}\mathcal{Q})
\end{eqnarray}
and again we obtain zero. This gives a consistency check on the
calculations of the NP constants using the definition in
(\ref{npconst}). In principle, higher multipole moments  of the Kerr
metric may also be obtained similarly at the cost of more complex
calculations of the higher order terms of $\Psi_0$.

\subsection{Physical Implications.}

The vanishing of the NP constants of the Kerr metric is puzzling in
connection with the no hair theorem for a black hole \cite{hair}.
The no hair theorem asserts that a Kerr black hole is  the unique
final state for gravitational collapse, like for instance in the
merger of  binary black holes with non-zero residual angular
momentum. Certainly we do not expect a generic
 initial data set which lead to eventual gravitational collapse
  will have vanishing NP constants
(see for instance \cite{DK}). But then how do we reconcile this with
the vanishing of the NP constants for the Kerr metric?

One way out, without compromising the no hair theorem,
 is that at the initial stage of a  black hole evolution,
the structure of null infinity is not smooth enough, i.e. the
conformal completion of the physical spacetime is $C^k,k<5$. The NP
constants are then not well defined at this early time of the
evolution. When the evolution enters a stage in which the Weyl
curvature falls off sufficiently rapid so that $\mathscr{I}^+$
becomes smooth enough, the NP constants begin to set in. Another
possibility we  should not overlook is that perhaps some hypotheses
of the no hair theorem may not be applicable for a generic collapse
situation.
 Certainly there are other possibilities we may think of.
The vanishing of the NP constants for the Kerr metric together with
the no hair theorem set a very stringent constraint for black hole
evolution. It is also worth understanding better to what extent  the
NP constants constrain
  the dynamics of gravitational collapse.

\section{Concluding Remarks}

One obvious shortcoming of the present work is that the construction
is valid only in a neighbourhood of null infinity. Unless we have an
analytic solution to the eikonal equation which matches to that
given here near null infinity, it is difficult to extend the
asymptotic coordinates  to cover entirely the Kerr metric exterior
to the event horizon. Still we hope the present work will provide a
small step towards our understanding of the gravitational waveform
of a relativistic rotating source. Further, the vanishing of the NP
constants of the Kerr metric also requires better understanding from
the perspective of black hole evolution and  gravitational wave
physics of a spacetime. This will hopefully enables us to gain
deeper insight into
   the physical meaning of these mysterious constants.

\section*{Acknowledgments} YKL would like to thank  Robert Bartnik
for some very useful discussions and   Helmut Friedrich for the
hospitality at AEI where part of this work was done. We enjoyed many
useful discussions with Sean Hayward during the course of the work.
We are grateful to Stephen Fletcher for sending us his thesis. The
work is supported by the NSF, China under contract number 10231050
and  by NKBRPC (2006CB805905).

\end{document}